\documentclass[aps,prb,twocolumn,superscriptaddress]{revtex4-1}

\usepackage{amsmath,amssymb}
\usepackage{graphicx}
\usepackage{color}
\usepackage{hyperref}
\usepackage{natbib}

\begin{document}

\title{The Nature of the Interlayer Interaction in Bulk and Few-Layer Phosphorus}
\author{L. Shulenburger}
\author{A.D. Baczewski}
\affiliation{Sandia National Laboratories, Albuquerque NM 87185}
\author{Z. Zhu}
\author{J. Guan}
\author{D. Tom{\'a}nek}
\affiliation{Michigan State University, East Lansing MI 48825}

\begin{abstract}
An outstanding challenge of theoretical electronic structure is 
the description of van der Waals (vdW) interactions in molecules and solids. 
Renewed interest in resolving this is in part motivated by the technological 
promise of layered systems including graphite, transition metal dichalcogenides, 
and more recently, black phosphorus, in which the interlayer interaction
is widely believed to be dominated by these types of forces. We report 
a series of quantum Monte Carlo (QMC) calculations for bulk black phosphorus 
and related few-layer phosphorene, which elucidate the nature of the forces 
that bind these systems and provide benchmark data for the energetics of 
these systems. We find a significant charge redistribution due 
to the interaction between electrons on adjacent layers. Comparison to density
functional theory (DFT) calculations indicate not only wide variability even among
different vdW corrected functionals, but the failure of these 
functionals to capture the trend of reorganization predicted by QMC. The delicate 
interplay of steric and dispersive forces between layers indicate that few-layer 
phosphorene presents an unexpected challenge for the development of vdW 
corrected DFT. 
\end{abstract} 

\maketitle

There is growing interest in understanding and correctly
describing the nature of the weak interlayer interaction in
layered systems ranging from few-layer graphene\cite{Novoselov2D}
to transition metal dichalcogenides\cite{Hennig1} such as MoS$_2$
and few-layer
phosphorene\cite{{Liu_ACSNano_2014},{Li_Nature_2014}}, which
display unique electronic properties and bear promise for device
applications. Anticipating that challenges concerning the
stability and isolation of single- to few-layer phosphorene can be
overcome\cite{stability-phosphorene}, this system with its unique
electronic\cite{{Liu_ACSNano_2014},{Li_Nature_2014}} and
optical\cite{Tran2014} properties is attracting particular
interest. It displays a high and anisotropic carrier
mobility\cite{{Lau-P2DM15},{Fei2014}} and a robust band gap that
depends sensitively on the in-layer strain\cite{Liu_ACSNano_2014}.
Progress in device fabrication\cite{{Koenig2014},{Li_Nature_2014}}
indicates clearly that phosphorene holds technological promise.
Since the fundamental band gap depends sensitively on the number
of layers\cite{Liu_ACSNano_2014}, understanding the nature of the
interlayer interaction is particulary important.

\begin{figure}[h]
 \centering
 \includegraphics[scale=0.35]{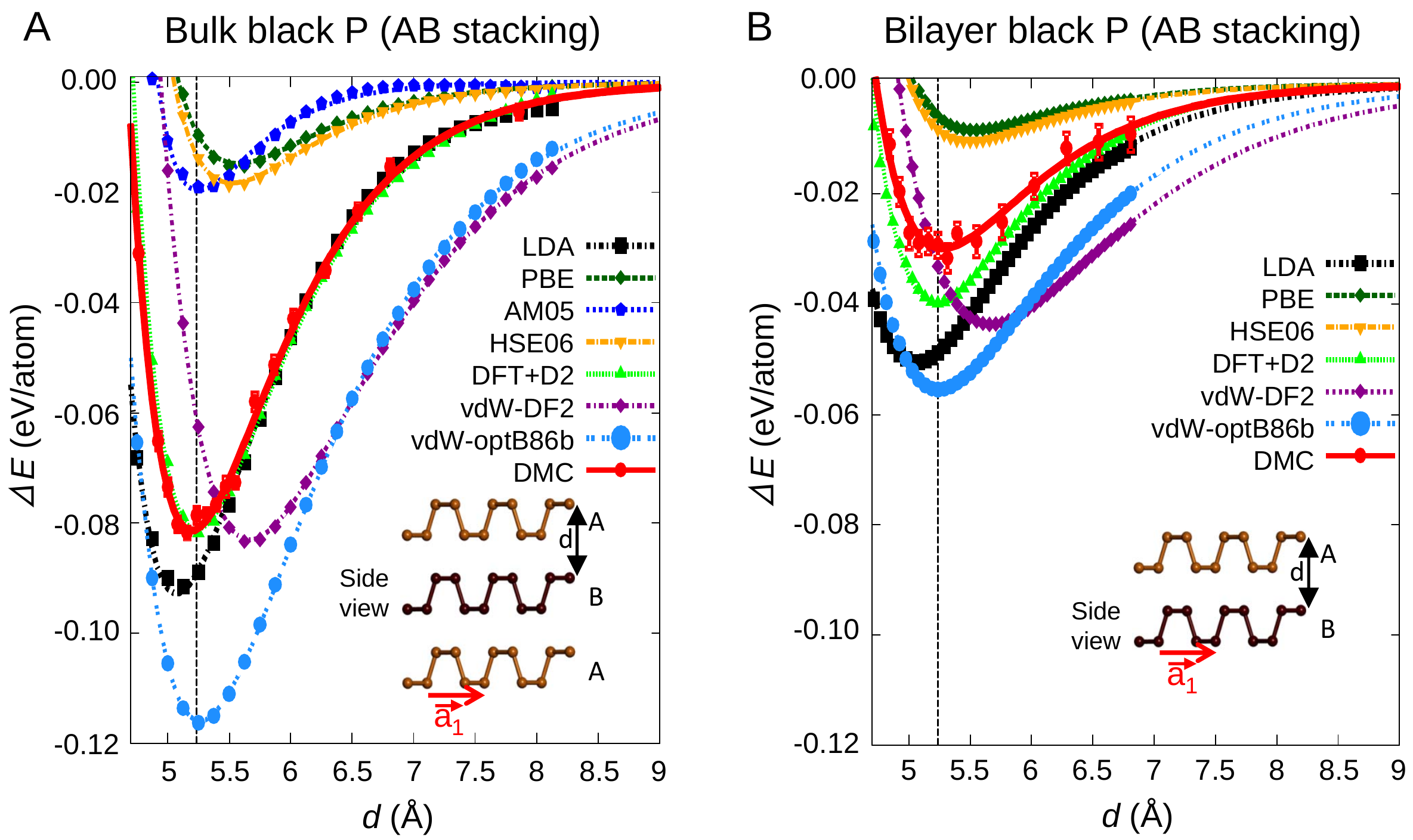}
 \caption{Binding energy per atom 
  $\Delta E$ as a function of the interlayer spacing $d$ in AB stacked 
  (A) bulk and (B) bilayer black phosphorus. QMC results obtained using
  diffusion Monte Carlo (DMC) are compared to DFT with
  LDA\cite{Perdew1981}, PBE\cite{Perdew1996}, AM05\cite{AM05},
  DFT+D2\cite{Grimme-D2}, vdW-DF2\cite{vdW-DF2},
  vdW-optB86b\cite{optB86b1,optB86b2} and HSE06\cite{HSE06}
  exchange-correlation functionals. The lines connecting the data
  points are Morse fits that extrapolate to ${\Delta}E=0$ for
  $d{\rightarrow}\infty$. The vertical dashed line indicates the
  observed interlayer spacing $d_e(expt.)=5.2365$~{\AA} in the bulk
  structure in (A) and the optimum value based on DMC in the bilayer
  in (B). Side views of the geometries are shown in the insets. \label{fig:fig1}}
\end{figure}

The standard approach to describe the interlayer interaction in
layered solids has been based on DFT. Whereas DFT is -- in
principle -- capable of describing the total energy of any system
in the ground state exactly, current implementations describe the
effects of electron exchange and correlation only in an
approximate manner. In most covalent and ionic solids of interest,
the specific treatment of exchange and correlation of electrons
does not play a crucial role and commonly used local or semi-local
exchange-correlation functionals of the electron density are
adequate. This approach may, however, not be adequate in complex
and weakly bonded systems\cite{Gygi2008} including black
phosphorus\cite{Appalakondaiah2012}. This is illustrated in
Fig.~\ref{fig:fig1}, which displays large differences between
interlayer interaction energies in bulk and bilayer black
phosphorus, obtained using different DFT functionals contained in
the {\sc VASP} software
package\cite{Kresse1996,Kresse1996a,Kresse1999}. The large spread
of the interlayer energies predicted using these functionals, 
some of which include van der Waals (vdW) corrections,
illustrates the gravity of the issue.

Even though vdW corrected exchange-correlation
functionals can improve the predicted geometry of layered
phosphorus, other quantities such as the transition pressure from
bulk orthorhombic to rhombohedral phases are significantly
underestimated\cite{Appalakondaiah2012}. This comes as no
surprise, since recent benchmarks in dense hydrogen indicate that
variations in the relative accuracy are observed among numerous
vdW corrected functionals depending on the quantity
being calculated\cite{Clay2014}. Even though the nature of the
interactions in phosphorus is quite different from hydrogen, the
difficulty to properly balance vdW and other
interactions likely persists.

A superior way to obtain insight into the nature of the interlayer
interaction requires a computational approach that treats electron
exchange and correlation adequately and on the same footing as
covalent and ionic interactions. Unlike in DFT, where the
fundamental quantity is the electron density, the fundamental
quantity in QMC calculations is the properly antisymmetrized
all-valence-electron wavefunction that explicitly describes the
correlation of electrons. Therefore, the weak interlayer
interaction in layered systems obtained using QMC is expected to
be more precise than the DFT-based counterpart, and the electron
density obtained by QMC likely provides a better representation of
the true charge density than DFT. Should the QMC-based electron
distribution in a few-layer system differ in a significant and
nontrivial manner from a superposition of electron densities in
isolated monolayers, we may surely conclude that the interlayer
interaction in such a system is not purely dispersive in nature.

In the last decade, QMC methods have demonstrated considerable
promise as a high accuracy first principles method for studying
solids\cite{Shulenburger2013} and are especially well-suited to
vdW bound systems
\cite{Drummond2006,Sorella2007,Spanu2009,Benali2014,Ganesh2014,%
Dubecky2013,Dubecky2014}. On one hand, the absence of approximations 
in treating the interaction between electrons makes it an ideal 
method for layered materials with a competition between different types of
interactions. On the other hand, the computational cost associated
QMC calculations is typically $1,000-10,000$ times higher than
that of comparable semilocal DFT calculations. Even though this
cost is mitigated by its superior parallel scalability, QMC is --
for the time being -- most useful for benchmark calculations.

We have performed QMC calculations using the diffusion Monte Carlo
(DMC) approach, as described in the Supplemental Material, in 
which details of the pseudopotential and an intensive procedure for 
converging finite size effects are included. One of the key approximations
in our calculation is bias due to a fixed nodal surface. This was not 
anticipated to be significant in black phosphorus as the binding of interest
occurs in a region of low electronic density for which the degree of nodal
nonlinearity is expected to be low \cite{Rasch2014}. 
Nevertheless, some DMC calculations were
carried out using orbitals from LDA, PBE, and vdW-optb86B functionals, to 
investigate the impact of the fixed node approximation. As the nodal 
surface associated with LDA orbitals were found to give the lowest energy
and our method is variational, the LDA orbitals were subsequently used in 
all cases.

Our DMC results for the interlayer binding curves in bulk and bilayer black
phosphorus are shown by the solid lines in Fig.~\ref{fig:fig1}.
The total energy difference between the monolayer and the bulk
system indicates that the binding energy of phosphorene sheets in
black phosphorus is $81{\pm}6$~meV/atom, which translates to a
cleavage energy of $22.4{\pm}1.6$~meV per {\AA}$^2$ of the
interface area. This is larger than that for many other layered
materials~\cite{Layered-Bonding}, but weak enough to allow
mechanical exfoliation that has been
reported.\cite{{Liu_ACSNano_2014},{Li_Nature_2014}} 

Comparing the DMC results to a variety of different DFT functionals, 
significant variability is evident.
Results are in agreement with intuition for LDA, PBE, and HSE06 
which do not treat vdW explicitly - namely that LDA  overbinds and
 reduces the interlayer spacing in comparison to experiment, 
 whereas GGA and HSE06 underbind.  The gradient corrected AM05, 
 designed to properly capture the energetics of the Airy gas at 
 a jellium surface, has none of the spurious self-interaction in 
 low density regions that cause LDA to overbind materials with vdW interactions. 
Nevertheless, it reproduces the bulk interlayer spacing correctly, but 
with only a quarter of the interlayer interaction energy predicted by DMC.
That a vdW-free functional binds at all gives some indication that 
the character of the interlayer interaction in these systems is not 
strictly vdW.  

From the wide array of available vdW functionals, 
we have selected 3 exemplary of increasing levels of sophistication. 
DFT+D2 is based upon an empirical correction to the total energy in 
the form of a simple pair-wise interaction parameterized by atomic $C_6$ coefficients. vdW-optB86b and vdW-DF2 are more advanced, both based upon 
improvements to the non-local vdW-DF functional \cite{vdW-DF}. Significant
variability is evident in both energetics and the equilibrium interlayer 
spacing among these functionals. It is interesting to note
that the least sophisticated of these functionals (DFT+D2) performs
the best relative to DMC, as well as experiment in the case of the
bulk system. 

Contrasting the bulk and bilayer binding curves, it is evident that 
DMC predicts that the interlayer interaction is not strictly additive.
For an additive interaction, we should expect that the binding energy of
the bilayer would be approximately 1/2 that of the bulk system because
both layers are missing half of their neighboring layers. Instead, DMC
predicts that this is not the case and that the bilayer binding energy 
is instead 3/8 that of the bulk system. The results for vdW corrected 
DFT (DFT+D2, vdW-opB86b, and vdW-DF2) are more indicative of an additive
interlayer interaction, giving us another indication that the nature
of the interlayer binding in black phosphorus is richer than a simple 
vdW interaction.

\begin{figure}[h]
\centering
\includegraphics[width=1.0\columnwidth]{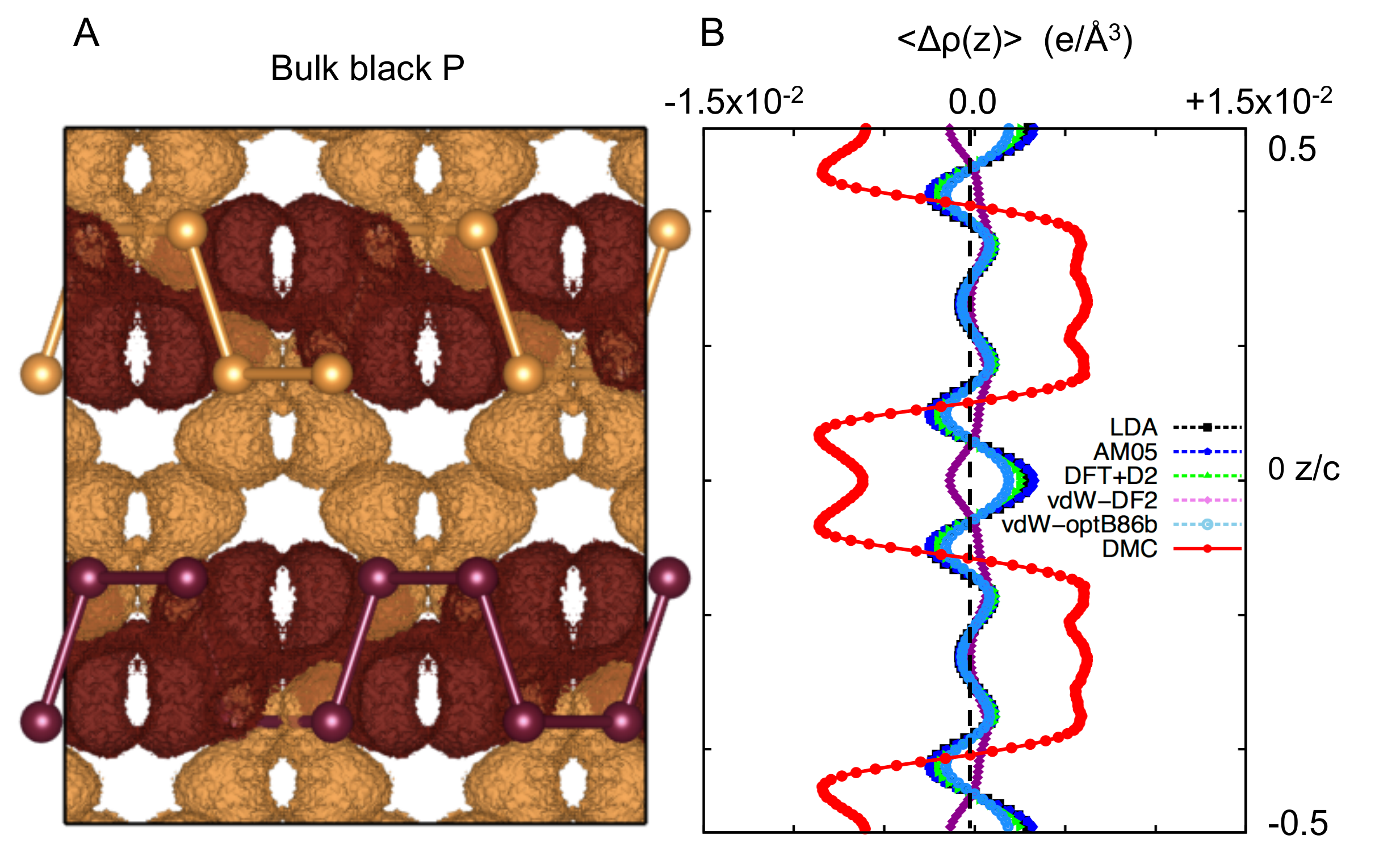}
\caption{Electron density difference $\Delta\rho = \rho_{tot}(bulk) - \sum \rho_{tot}(monolayers)$ representing the charge redistribution caused by assembling the bulk structure from isolated monolayers. (A) DMC isosurfaces bounding regions of excess electron density (dark brown) and electron deficiency (light brown), with respective values $\pm 6.5\times10^{-3} \text{e}/\AA^3$. (B) $\langle \Delta \rho(z) \rangle$ for DMC and several DFT functionals averaged across the $x-y$ plane of the layers, with $z/c$ indicating the relative position of the plane in the unit cell. 
\label{fig:fig2}}
\end{figure}

A more direct indication of the character of the
interlayer binding is the charge density difference induced by
assembling the bulk system from isolated monolayers. We computed the 
quantity ${\Delta}\rho = \rho_{tot}(bulk)-\sum\rho_{tot}(monolayers)$
using both DMC and DFT to investigate this. The $l_1$-norm of ${\Delta}\rho$ 
over the unit cell is an indicator of the number of electrons being 
redistributed due to interlayer interaction. In all DFT functionals considered
in this study, this metric indicates a motion of fewer than 0.03 electrons per
atom. In contrast, it predicts a motion of 0.15 electrons per atom in DMC.
To provide insight into the nature of this significant redistribution, 
the density difference is visualized in Fig.~\ref{fig:fig2}A. Inspection 
indicates that charge is pushed out of the
region between layers and into the covalent bonds within each layer. This 
picture is well supported by basic chemical intuition. In an isolated layer,
 each atom is 3-fold coordinated with $sp^3$ bonding character and a single 
 lone pair protruding away from the layer. Bringing layers together will 
 increase the overlap between these lone pairs on adjacent layers and steric 
 forces will tend to drive the affiliated lone pair charge closer to the 
 layer on which it originated. 
 
To further elucidate the role that this charge redistribution plays in the 
interlayer binding, the planar average of the charge density difference
along planes perpendicular to the interlayer axis for both DMC and several
 DFT functionals are illustrated in Fig.~\ref{fig:fig2}B. It remains evident 
 that DMC predicts an average depletion of charge between layers and commensurate
 accumulation in-layer.  However, with the exception of vdW-DF2, this trend is 
 not evident in any of the DFT calculations. Instead, DFT predicts a weak 
 accumulation of charge between layers, indicating a minor covalent character
 in addition to whatever vdW binding is present. This likely 
 explains why a vdW-free functional like AM05 weakly binds the system
near the correct interlayer spacing. The case of vdW-DF2 is of particular interest
because it is the only functional that predicts a qualitatively similar, but 
considerably weaker, trend as DMC. Among DFT+D2, vdW-optB86b, and vdW-DF2, vdW-DF2
is the most theoretically sophisticated so that it gets closer to the trend
predicted by DMC is unsurprising. Even so, that DFT+D2 gives the best performance
in terms of energetics and geometry indicates that this functional may be getting
the right answer for reasons that are not necessarily consistent with the 
many-body physics more explicitly explored through DMC.

Recent work on self-consistent vdW functionals indicate that the charge 
redistribution induced by vdW interactions can play an important role in the 
energetics of highly polarizable systems \cite{Ferri2015}. In this work, the
authors note that this subtle physics is consistent with an early observation
by Feynman \cite{Feynman39} in which the vdW interaction can be viewed as 
arising from an attractive interaction induced through a small accumulation 
of charge density between two mutually perturbed neutral systems. In the case
of black phosphorus, we find that this picture of the vdW interaction is balanced 
by the steric redistribution of charge away from the region between layers. Based
upon the results elucidated by our DMC calculations, we anticipate that getting 
this balance right may be a critical requirement to examine in developing more 
advanced DFT functionals for layered compounds.

\begin{figure}[h]
\centering
\includegraphics[width=0.75\columnwidth]{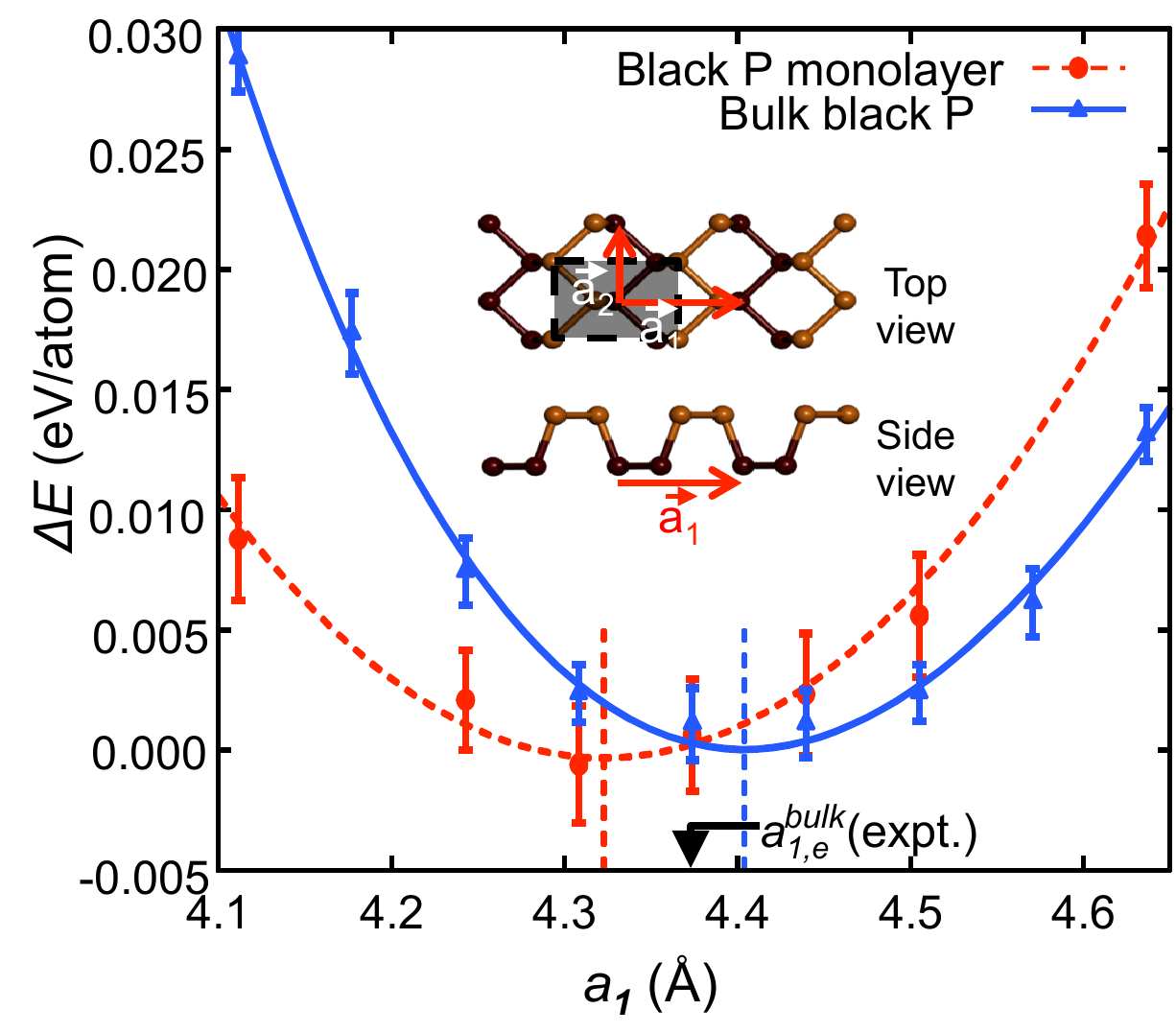}
 \caption{DMC results for the relative total energy per atom ${\Delta}E$
 as a function of the in-layer lattice constant $a_1$ in the soft
 direction of a phosphorene monolayer and of bulk black phosphorus.
 The lines connecting the data points are Morse fits that
 extrapolate to ${\Delta}E=0$ for $d{\rightarrow}\infty$. The
 optimum lattice constant values for both structures are indicated
 by the vertical dashed lines. The observed value
 $a_{1,e}(expt.)=4.374$~{\AA} in the bulk structure is indicated by
 the arrow. The monolayer geometry is shown in the
 inset.\label{fig:fig3}}
\end{figure}

The corresponding difference between the charge density in an
isolated monolayer and within a layer of bulk phosphorus is likely
to affect the in-plane bonding and geometry. To see if this is
indeed the case, we have calculated the energy change ${\Delta}E$
as a function of a stretch applied along the softer axis
$\vec{a}_1$ of the sheets. Our results for the bulk system,
presented in Fig.~\ref{fig:fig3}, indicate an excellent
agreement (to within half a percent) between the optimized lattice
constant $a_1(theory)=4.404{\pm}0.019$~{\AA} and the observed
value~\cite{Cartz1979} $a_1(expt.)=4.374$~{\AA} in the bulk
structure. Further, we can see precisely how soft this axis
is in both systems, with a deformation by $|{\Delta}a_1|{\lesssim}0.3$~{\AA} 
requiring an energy investment of only ${\approx}5$~meV/atom in a 
monolayer or in bulk black phosphorus. This energy corresponds to 
a thermal energy of $60$~K, and we should expect significant
thermal fluctuations of the geometry of unsupported phosphorene
sheets at ambient temperature and pressure.
Most important, however, is the comparison between $a_1$ in
the isolated monolayer and in the bulk structure. Our DMC results
indicate a change in the in-plane stiffness along the soft axis
and an ${\approx}2$\% reduction in the equilibrium lattice
constant $a_1$ in a monolayer from the bulk value. This is another
indication of a charge redistribution during the formation of a
layered bulk structure from monolayers that modifies the covalent
interaction within the layers. This again supports our finding
that the interlayer interaction in black phosphorus is not purely
dispersive.

\begin{figure}[h]
\centering
\includegraphics[width=1.0\columnwidth]{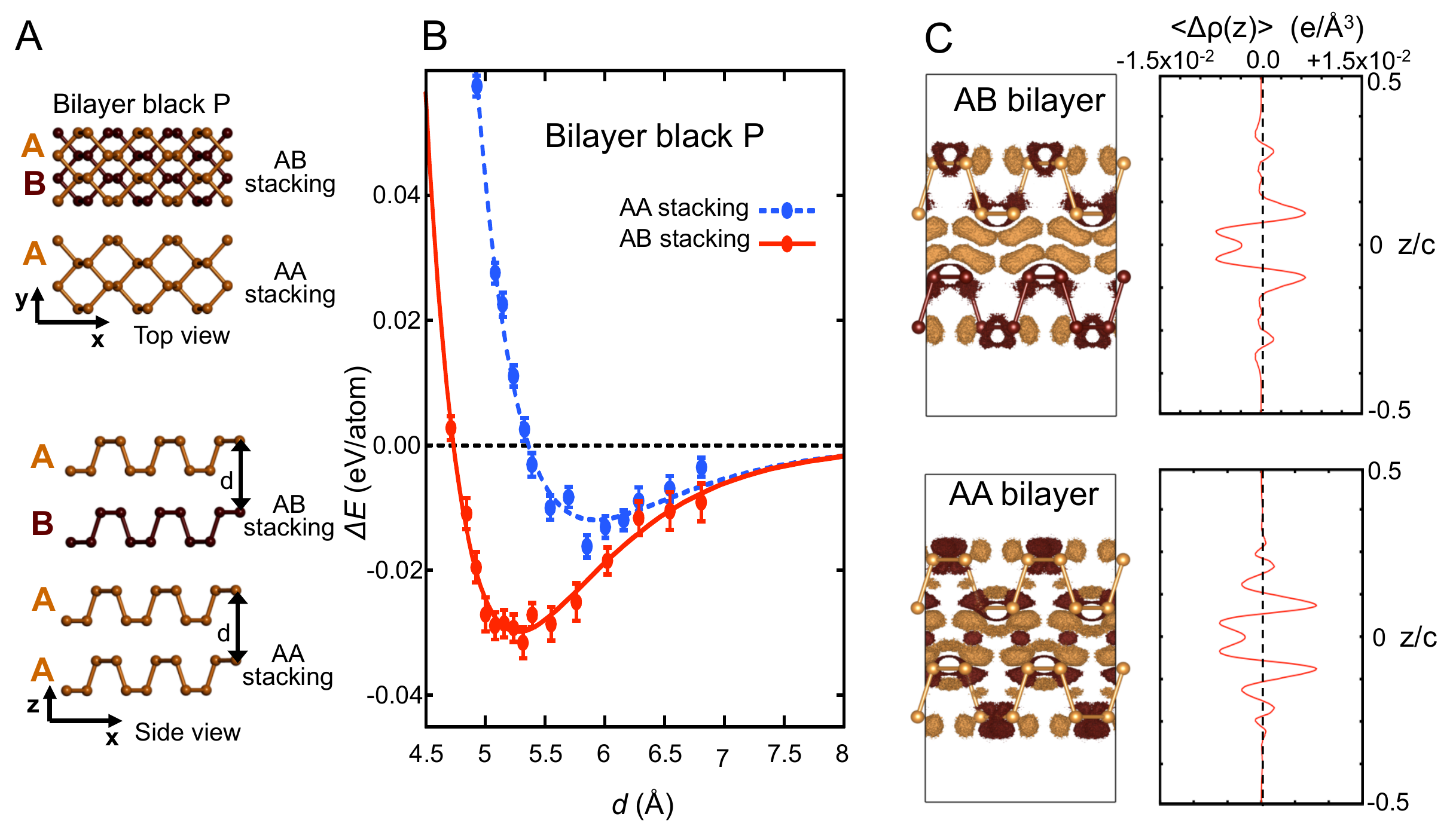}
\caption{DMC results for bonding in AA and AB stacked phosphorene bilayers. (A) Geometry of an AA and AB stacked bilayer in top and side view. (B) DMC results for the relative total energy per atom $\Delta E$ as a function of the interlayer spacing $d$. The lines connecting the data points are Morse fits that extrapolate to $\Delta E=0$ for $d \rightarrow \infty$. (C) Electron density differences for the AB and AA bilayers illustrated using isosurfaces and planar averaging as in Fig.~\ref{fig:fig3} (with the same color coding).
\label{fig:fig4}}
\end{figure}

To gain additional insight into the nature of the interlayer
interaction, we compare in Fig.~\ref{fig:fig4} the bonding within
an AA and AB stacked bilayer as a function of the interlayer
separation $d$. Our DMC results in Fig.~\ref{fig:fig4} indicate
that the AB stacking, which occurs in the bulk material, persists
also in the bilayer.

The cleavage energy of an AB stacked bilayer
is $16.6{\pm}2.2$~meV/{\AA}$^2$, thus 26\% smaller than the bulk
cleavage energy of $22.4{\pm}1.6$~meV/{\AA}$^2$. The exfoliation
energy associated with removing the topmost layer from the surface
is expected to lie between these two values. The findings appear
plausible also in view of the fact that in graphite, the cleavage
energy is estimated to be 18\% larger than the exfoliation
energy\cite{Girifalco1956}. Though we expect the interlayer
interaction to be mediated primarily by the $\pi$ electrons in
graphite and $sp^3$-like lone pairs with a different character in
black phosphorus, the ratio of the exfoliation and cleavage energy
in the two systems appears to be similar.

The interlayer spacing $d_e=5.272{\pm}0.023$~{\AA} in the
AB-stacked bilayer is about $1$\% larger than our calculated
value for the bulk material. In the less favorable AA stacking
geometry, the binding energy is three times smaller than in the AB
geometry, which should effectively prevent formation of stacking
faults, at least in the absence of impurities. The significant
difference between the interaction in the AA and AB stacked
bilayer suggests that the interaction between the sheets is more
complicated than the vdW interaction between two
homogeneous slabs. Were the interlayer interaction purely
dispersive, the registry of layers would not matter much and the
AA and AB binding energies as well as interlayer separations
should be nearly identical.

To shed some light on the sensitivity of the interlayer bonding on
the stacking sequence, we investigated the change in the charge
density ${\Delta}\rho$ induced by the interaction. Our results for
AB and AA stackigns are shown in Fig.~\ref{fig:fig4}\textit{C}. 
Similar to the corresponding results for bulk black phosphorus in
Fig.~\ref{fig:fig2}\textit{B}, the ${\Delta}\rho$ plots for the
bilayer show a significant rearrangement of the electronic charge,
a total of $\approx$0.075~electrons per atom in both cases. 
In the AB-stacked bilayer,
similar to the bulk system, we observe a depletion of the electron
density in the region between the sheets and electron accumulation
within the layers. The charge redistribution in the AA bilayer is
significantly different, even including small regions in the interlayer
space where the charge density increases. The large difference
between ${\Delta}\rho$ in the AA and AB stacked bilayer is
inconsistent with purely dispersive bonding and explains why the
the interlayer spacing and binding energy are so different in the
two systems.

In summary, we studied the nature of the interlayer interaction in
layered black phosphorus using quantum Monte Carlo calculations,
which describe the correlation of electrons explicitly by an
antisymmetrized all-valence-electron wavefunction and treat
covalent and dispersive interactions on the same footing. Unlike
in true vdW systems, we find that the interlayer
interaction in few-layer phosphorene is associated with a
significant charge redistribution between the in-layer and
interlayer region, caused by changes in the non-local correlation
of electrons in adjacent layers. Consequently, the resulting
interlayer interaction can not be described properly by density
functional theory (DFT) augmented by mere semi-local vdW
correction terms, and thus the designation `van der Waals solids'
is improper for systems including few-layer phosphorene. Our
results may be used as benchmarks for developing more
sophisticated DFT functionals that should provide an improved
description of non-local electron correlations in layered systems.

We are grateful for useful comments and conversations with Paul Kent,
Jeongnim Kim, Ann Mattsson, Jonathan Moussa, Lydia Nemec, 
and Gotthard Seifert. Calculations were performed on {\em
Sequoia} at Lawrence Livermore National Laboratory and {\em Mira}
at the Argonne Leadership Computing Facility. We thank Anouar
Benali for assistance performing calculations on {\em Mira}. An
award of computer time was provided by the Innovative and Novel
Computational Impact on Theory and Experiment (INCITE) program
with Project CPH103. A.B. and L.S. were supported through the
Predictive Theory and Modeling for Materials and Chemical Science
program by the Office of Basic Energy Sciences (BES), Department
of Energy (DOE). Z.Z., J.G., and D.T. acknowledge support by the
National Science Foundation Cooperative Agreement No. EEC-0832785,
title ``NSEC: Center for High-Rate Nanomanufacturing''. Sandia
National Laboratories is a multi-program laboratory managed and
operated by Sandia Corporation, a wholly owned subsidiary of
Lockheed Martin Corporation, for the U.S. Department of Energy's
National Nuclear Security Administration under contract
DE-AC04-94AL85000.

\pagebreak
\widetext

\begin{center}
\textbf{Supplemental Materials for: The Nature of the Interlayer Interaction in Bulk and Few-Layer Phosphorus}
\end{center}
\setcounter{equation}{0}
\setcounter{figure}{0}
\setcounter{table}{0}
\setcounter{page}{1}
\makeatletter
\renewcommand{\theequation}{S\arabic{equation}}
\renewcommand{\thefigure}{S\arabic{figure}}
\renewcommand{\bibnumfmt}[1]{[S#1]}
\renewcommand{\citenumfont}[1]{S#1}

\section{Convergence of Finite-Size Effects}

One- and two-body finite size effects must be considered in both the bulk and planar periodic structures. One-body effects are corrected using canonical twist averaging \cite{twist-averaging}, while two-body effects are  extrapolated. Controlling the two-body effects present a challenge for extrapolation consistent with reports of calculations done on graphite.\cite{Spanu_PRL_2009}

For the bulk systems, we consider two types of tilings for generating these supercells. We expect finite size effects to converge more quickly when increasing the effective system size along the layering axis rather than in-plane. Consequently, we study two-body effects for 2x2x1, 3x3x1, 4x4x1, and 5x5x1 tilings (single layer) and 2x2x2, 3x3x2, and 4xe4x2 (bilayer) in the bulk. 

However, the peculiar nature of the electron correlation in two dimensional systems such as black phosphorus coupled with the meV accuracy required for this study requires a slightly different procedure for assessing and reducing the two body finite size effects introduced by simulating a supercell with periodic boundary conditions.  The typically used procedures for this involve either utilizing a model interaction that removes spurious electron correlation,\cite{MPC} analyzing the behavior of the structure factor and two body jastrow factor for small values or momentum transfer\cite{Chiesa-finite-size} or utilizing calculations with density functionals designed to mimic the energetics of the electron correlation in finite supercells.\cite{kzk}  As generally practiced, all of these procedures rely on the assumption that the electron correlation is isotropic, a condition that is grossly violated in a layered material like black phosphorus.

\begin{figure}[ht]
\centering
\includegraphics[scale=0.35,angle=270]{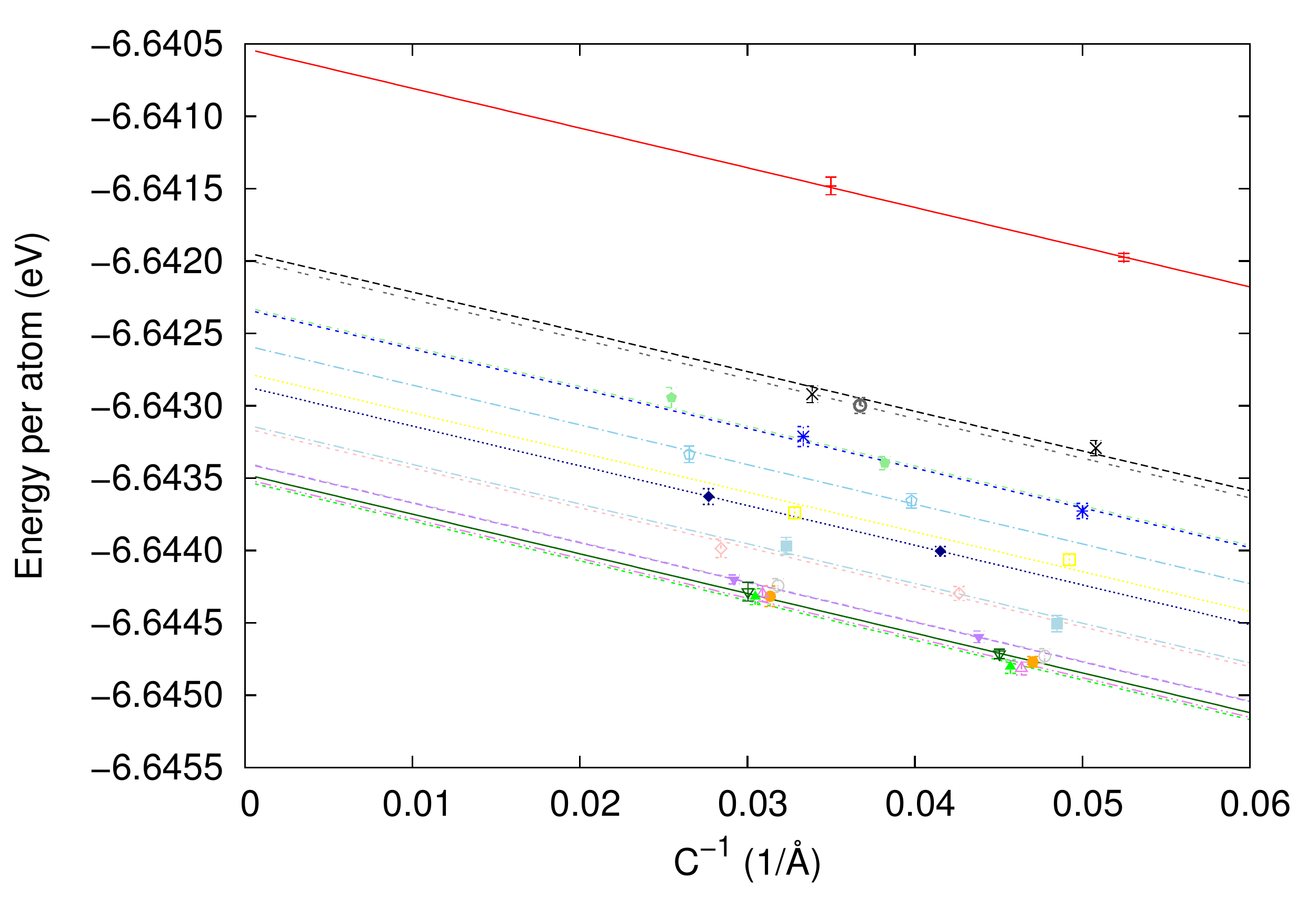}
\caption{Energy per atom of black phosphorus as a function of the inverse of the shortest distance between periodic images of phosphorene sheets.  The identical slope of the lines enforces consistency in the extrapolation for calculations where the distance between adjacent sheets are varied.\label{fig:extrapolation}}
\end{figure}

This difficulty has been noted in previous quantum Monte Carlo studies of layered materials\cite{Spanu_PRL_2009} where a combination of the KZK functional\cite{kzk} was used for correlations in the plane and an extrapolation scheme was used for correlations in the direction perpendicular to the planes.  Our approach is to extend this methodology in two regards.  Firstly, we utilize extrapolation in the size of the supercell to determine correlations both in plane and out of plane and secondly, because such extrapolations can introduce a significant amount of noise in the extrapolated quantities, we correlate the parameters of the extrapolations between systems with similar geometries.  This is best illustrated looking at the extrapolation of the energy in the direction perpendicular to the phosphorene planes.  

This correlation may be expected to decrease with distance between the planes and also be a function of the number of layers simulated.  To capture this, we extrapolate based on the total distance between images in the perpendicular direction and also require that the slope of the extrapolation be a simple function of the distance between the planes.  The results of this procedure are shown in fig~\ref{fig:extrapolation}, where the energy per atom for different supercells with different spacings between the phosphorene planes and different numbers of layers in the supercell are shown as a function of the inverse of the distance between the top and bottom of the supercell.  All of these calculations had a three by three tiling of the primitive cell of phosphorus in the direction parallel to the planes and both the spacing between the layers and the number of copies of the supercell were varied.  A similar procedure was used to extrapolate to infinite size supercells in the other direction.

\section{Numerical Considerations}

To control computational cost, we verify that our choice of DMC time step satisfies a balance between efficient sampling and having a large time step bias. In doing so, for each system we choose fixed moderately sized supercells and perform short DMC runs at different time steps. A best fit line is constructed for the energy as a function of time step and we choose to use the largest time step that is within 1 mHa/atom of the zero time step extrapolated energy.  We have found that a time step of 0.0075 a.u. is adequate to achieve this level of accuracy in all cases.

Further, to control the memory required by the wave function we assess the effect of enlarging the grid spacing associated with the B-spline representation of the Kohn-Sham orbitals relative to the equivalent real space grid used in the plane wave pseudopotential calculation in which they are generated. The convergence of the total energy, kinetic energy, and variance in the local energy are assessed in determining the appropriate grid spacing.  Noting that the variance converges most slowly in the grid spacing, an enlargement factor of 4/3 preserves accuracy while reducing the memory used up in representing the wave function. 

\section{Pseudopotential generation and testing}

Two different pseudopotentials were generated for this study, one treating 5 valence electrons and the other treating 13. Both pseudopotentials were generated using the opium pseudopotential generation code.\cite{opium} The benchmark quantities under consideration are the equilibrium properties of a phosphorus dimer and the ionization potential and electron affinity of an isolated phosphorus atom.  

In Table \ref{tab:atomic_benchmarks} some atomic benchmarks are given.  While the accuracy relative to experiment is comparable for the ionization potential for both pseudopotentials, the 5 electron pseudopotential performs worse for the electron affinity.  Given the relatively primitive trial wavefunction used, these results should not be taken as conclusive, but errors due to the nodal surface are generally less than a few tenths of an eV.

\begin{table}[ht]
\begin{center}
    \begin{tabular}{ | c | c | c | c |}
    \hline
     & 5 electron & 13 electron & Experiment \\ \hline
     Ionization Potential & 10.7112 $\pm$ 0.00084 & 10.6832 $\pm$ 0.0598 & 10.48669 \\ \hline
     Electron Affinity & 0.6405 $\pm$ 0.0084 & 0.7483 $\pm$ 0.0626 & 0.746609 \\ \hline
    \end{tabular}
\end{center}
\caption{Ionization potential and electron affinity of an isolated phosphorus atom computed using two different pseudopotentials. \label{tab:atomic_benchmarks}}
\end{table}

\begin{figure}[ht]
 \centering
 \includegraphics[scale=0.35,angle=270]{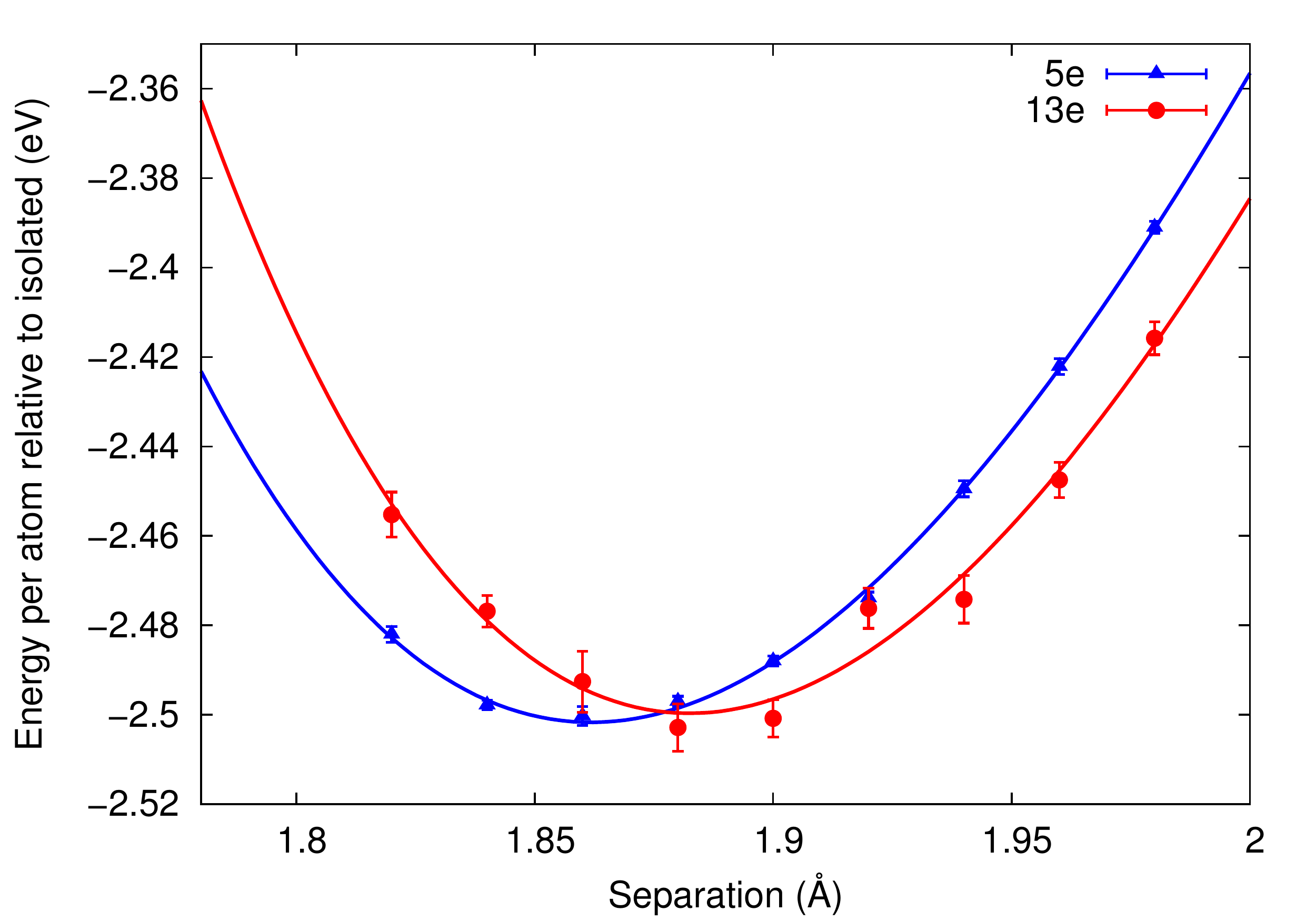}
 \caption{Phosphorus dimer energy vs bond length curves for 5 and 13 electron pseudopotentials. Experimental values for the ionization potential and electron affinity are taken from References \onlinecite{0953-4075-44-19-195009} and \onlinecite{1402-4896-22-3-010} respectively. \label{fig:dimer_benchmarks}}
\end{figure}

We also calculate the binding energy, atomization energy and vibrational frequency for the phosphorus dimer by calculating the energy as a function of bond length.  The results of these calculations for each pseudopotential and a fit to a morse potential are shown in Figure \ref{fig:dimer_benchmarks} and the resulting observables are summarized in Table \ref{tab:dimer_benchmarks}. 

\begin{table}[ht]
\begin{center}
    \begin{tabular}{ | c | c | c | c |}
    \hline
     & 5 electron & 13 electron & Experiment \\ \hline
     bond length (\AA) & 1.8618 $\pm$ 0.009 & 1.8824 $\pm$ 0.0018 & 1.89340 $\pm$ 0.00044 \\ \hline
     vibrational frequency (cm$^
{-1}$)& 827.3 $\pm$ 9.4 & 851.7 $\pm$ 31.8 & 780.77 \\ \hline
     atomization energy (eV) & 4.900 $\pm$ 0.001 & 4.894 $\pm$ 0.040 & 5.03 $\pm$ 0.02 \\ \hline
    \end{tabular}
\end{center}
\caption{Properties of phosphorus dimer computed using two different pseudopotentials.  The experimental bond length, vibrational frequency and atomization energy are taken from references \onlinecite{huber1979constants}, \onlinecite{p2-vibrational} and \onlinecite{huber1979constants} respectively. \label{tab:dimer_benchmarks}}
\end{table}

While both the 5 electron and 13 electron pseudopotentials have similar accuracy relative to experiment, calculations using the potential with 13 electrons in the valence are nearly 100 times more expensive than those using the 5 electron potential.  Therefore, its success in these metrics leads us to use the 5 electron pseudopotential for all remaining calculations on bulk phosphorus and phosphorene.

\end{document}